\documentstyle[twoside,fleqn,npb,epsfig,axodraw]{article}
\input{diagrams.axd}

\title{Mathematics for structure functions \hfill NIKHEF-00-008}
\author{J.A.M. Vermaseren and S. Moch\address{
 NIKHEF Theory Group\\
        Kruislaan 409, 1098 SJ Amsterdam, The Netherlands}}
\begin{document}

\maketitle

\section{Introduction}
Over the years a particular type of mathematics has been developed for the 
evaluation of the integrals that occur in Feynman diagrams with loops. The 
solution of these integrals often involves the introduction of Feynman 
parameters and the manipulation of a special set of functions. Yet we are 
running into limitations imposed by the complexity of the problem. Hence 
most quantities that are important for comparing theoretical predictions 
and experimental observations have not been evaluated beyond two loops. Yet 
it is important that a number of these quantities will be computed at the 
three loop level. Among these are the QCD structure functions in deep 
inelastic scattering. Considering the accurate measurements at HERA and the 
need for precise parton distribution functions at the LHC this may 
currently be considered as one of the most important calculations to be 
done. The conventional techniques have thus far not led to any results. 
Hence the introduction of different techniques may be called for. One 
method for the evaluation of structure functions has been around from the 
beginning of QCD~\cite{Gross:1973rr}\cite{Politzer:1974}, but has been used 
only very occasionally and then not to the fullest of its potential. This 
method is the evaluation of the anomalous dimensions and/or the coefficient 
functions in Mellin space in such a way that all even or all odd moments 
are determined. When all even or all odd Mellin moments of a function are 
known the function of which these are the Mellin moments is uniquely 
determined~\cite{Santiago:1999pr}. This is within a resonable set of 
assumptions about behaviour of the functions involved. Hence a systematic 
method to obtain these Mellin moments would be sufficient to obtain the 
anomalous dimensions and/or the coefficient functions. This method has been 
applied several times in the 
past~\cite{Gonzalez-Arroyo:1979df}\cite{Kazakov:1988jk}\cite{Moch:1999c}, 
but the original 
computation~\cite{vanNeerven:1991nn} of the full set of two loop 
coefficient functions used a more conventional method. Here I will report 
on attempts to use this Mellin method even at the three loop level. This 
needs a very careful study of the functions involved. Much of this is new 
mathematics. We will address here harmonic sums, harmonic polylogarithms, 
Mellin transforms, inverse Mellin transforms and difference equations.

\section{Harmonic sums}

The basic functions that occur in Mellin space are the harmonic sums 
(HS)~\cite{Knuth}\cite{Vermaseren:1998uu}\cite{Blumlein:1998if}. 
These are defined by
\begin{eqnarray}
\label{eq:basicharmo}
S_m(N) = \sum\limits_{i=1}^N  \frac{1}{i^m}\, , \quad 
S_{-m}(N) = \sum\limits_{i=1}^N  \frac{(-1)^m}{i^m}\, ,
\end{eqnarray}
while `higher' sums can be defined recursively
\begin{eqnarray}
S_{m_1,...,m_k}(N) &=& 
        \sum\limits_{i=1}^N  \frac{1}{i^{m_1}} S_{m_2,...,m_k}(i)\, , \\
S_{-m_1,...,m_k}(N) &=& 
        \sum\limits_{i=1}^N  \frac{(-1)^{m_1}}{i^{m_1}} S_{m_2,...,m_k}(i)\, .
\end{eqnarray}
The parameters $m$ are called the indices, while $N$ is called the 
argument. We define the weight of a sum by the sum of the absolute values 
of its indices. Whereas the two loop structure functions needed only sums 
of up to weight 4~\cite{Vermaseren:1998z}, we expect the three loop 
coefficient functions to need sums of weight 6.
There exists an alternative notation in which the indices take only the 
values 1,0 and -1. In this notation a 0 indicates that the nearest nonzero 
index to the right of this zero should be raised by one in its absolute 
value. Hence
\begin{equation}
	S_{0,1,0,0,-1}(N)  =  S_{2,-3}(N)
	 =  \sum\limits_{i=1}^n \sum\limits_{j=1}^i 
			\frac{\sign(j)}{i^2j^3}
\end{equation}
In this notation one can see easily how many independent sums there are for 
a given weight. All functions of a given weight $w$ have exactly 
$w$ indices. All these indices can have the values 1,0,-1 with the 
exception of the last one, which can have only the values 1 and -1. Hence 
there are $2\ 3^{w-1}$ independent sums of weight $w$. One of the nice 
properties of these sums is that they form an algebra. This means that the 
product of two sums with the weights $w_1$ and $w_2$ respectively can be 
expressed as a sum of terms, each with a single harmonic sum of weight $w 
= w_1+w_2$.

These sums show themselves in a variety of ways. The most immediate is the 
expansion of the $\Gamma$-function.
\begin{eqnarray}
	\Gamma(-n+\epsilon) & = & \frac{\sign(n)}{\epsilon n!}
		\Gamma(1+\epsilon)(1+S_1(n)\epsilon \nonumber \\ &&
	+S_{1,1}(n)\epsilon^2
	+ S_{1,1,1}(n)\epsilon^3  \nonumber \\ &&
	+ S_{1,1,1,1}(n)\epsilon^4
	+ \cdots )
\end{eqnarray}
At the positive side one has a slightly messier expansion.

Another one is in sums of the type
\begin{eqnarray}
        \sum_{i=1}^n \sign(i) \binom(n,i) \frac{1}{n^3} & = &
                        -S_{1,1,1}(n)\, ,
\end{eqnarray}
In general we will need sums over these sums. This can be set up in a 
variety of ways. In some ways the multiple sums are so intertwined that we 
cannot resolve them. In other ways we can arrange that we have always one 
parameter sums, meaning that for instance the upper limit of the summation 
and the summation parameter itself are the only parameters in the problem 
and hence the upper limit is the only parameter in the answer. If our sums 
can be expressed as nested one-parameter sums we make a good chance to be 
able to resolve them. Some examples of such single parameter sums that can 
be handled are~\cite{Vermaseren:1998uu}:
\begin{eqnarray}
&\ \ \ \ & \sum_{i=1}^n \frac{S_{\vec{p}}(i)S_{\vec{q}}(n\minus i)}{i^m}, \nonumber \\
&\ \ \ \ & \sum_{i=1}^n \sign(i)\frac{S_{\vec{p}}(i)S_{\vec{q}}(n\minus i)}{i^m},\nonumber \\
&\ \ \ \ & \sum_{i=1}^n \sign(i)\binom(n,i)\frac{S_{\vec{p}}(i)}{i^m}, \nonumber \\
&\ \ \ \ & \sum_{i=1}^n \sign(i)\binom(n,i)\frac{S_{\vec{p}}(i)S_{\vec{q}}(n\minus i)}{i^m}.
                \nonumber
\end{eqnarray}
If we assume that each diagram can be expressed in terms of harmonic sums 
and we can bring a hierarchy amoung the diagrams in which the more 
complicated diagrams can be expressed as single sums over simpler diagrams, 
then the above condition is fulfilled.

\section{Harmonic polylogarithms}

In the end we would like to have functions of $x=Q\mydot Q/2P\mydot Q$ 
rather than just the Mellin moments. When the two-loop structure functions 
are computed directly as functions of $x$, one runs into complicated 
integrals of which the solutions can be expressed in terms of dilogarithms 
and trilogarithms~\cite{Lewin} of composite arguments. At the three loop 
level one could imagine that tetra- and penta-logarithms might do the job, 
but it turns out that these functions are not sufficient. Also the Nielsen 
polylogarithms~\cite{Nielsen} are not sufficient. Hence we need the so 
called harmonic polylogarithms (HP)~\cite{Remiddi:1999ew} which are also 
closely related to multi dimensional polylogarithms~\cite{Broadhurst:1998}.
\begin{eqnarray} 
  H(0;x) &=& \ln{x} \ ,          \nonumber\\ 
  H(1;x) &=& \int_0^x \frac{dx'}{1-x'} = - \ln(1-x) \ , \nonumber\\ 
  H(-1;x) &=& \int_0^x \frac{dx'}{1+x'} = \ln(1+x) \ , \nonumber \\
  H(\vec{m}_w;x) & = &
			 \int_0^x dx' \ f(m_w;x') \ H(\vec{m}_{w-1};x') \ , 
		\nonumber\\
   f(0;x) &=& \frac{1}{x} \ , \nonumber\\
   f(1;x) &=& \frac{1}{1-x} \ , \nonumber\\
   f(-1;x) &=& \frac{1}{1+x} \ , \nonumber \\
  H(\vec{0}_w;x) & = & \frac{1}{w!} \ln^w{x} \ 
\end{eqnarray}
in which $\vec{m}_w$ stands for $m_w,\cdots,m_1$. 
One can derive a number of transformations for these functions that allow a 
moderately easy evaluation of these functions in the entire complex plane 
with the exception of the points around $+i$ and $-i$ where convergence is 
rather slow. We will need them on the interval $0-1$ only.
When one makes a taylor expansion of these functions, the coefficients are 
combinations of harmonic sums.

One can define weights for these functions in the same way as was done for 
the harmonic sums. It shows that for each weight there are $3^w$ different 
harmonic polylogarithms. Like the harmonic sums the harmonic polylogarithms 
form an algebra, although the rules are somewhat different. This makes it 
very easy to bring expressions into an unique standard form.

\section{Mellin transforms}

The Mellin transform is defined by
\begin{equation}
	M(f(x)) = \int_0^1dx\ x^{N-1} f(x)
\end{equation}
It is not significant whether we use $N-1$ or $N-2$ in this definition. 
When we do such integrals the result can be expressed in a combination of 
harmonic sums in $N$ and harmonic sums in infinity. The latter ones are just 
constants. Some efforts have been spent recently on reducing these 
constants to a minimal set. This has been programmed up to weight 
9~\cite{Broadhurst:1997}\cite{Vermaseren:1998uu}, which 
should be sufficient even for the five loop anomalous dimensions.
We will need them to weight 6 in which case there are `only' 8 different 
constants and combinations of them.

The interesting part comes when we look at Mellin transforms of functions 
of the types $H(x)/(1-x)$ and $H(x)/(1+x)$.
For a given weight $m$ there are $3^m$ functions $H$ and hence the above 
set of Mellin transforms has $2\ 3^m$ elements, which is identical to the 
number of harmonic sums of weight $m+1$. Indeed there is a one to one 
correspondence between the above Mellin integrals and harmonic sums of 
weight $m+1$. This is to say that each of these integrals results in a 
single harmonic sum of weight $m+1$ and a number of harmonic sums of lower 
weight, possibly multiplied by some of these constants.
This allows for a method of inverse Mellin transformation, but there is 
still one little problem. When using the Mellin transform one obtains 
not only harmonic sums but there may also be a factor $\sign(N)$. This 
doubles the number of potential terms in the space of harmonic sums. Let us 
write it differently however. Instead of using the factors $1$ and $\sign(N)$ 
we use the factors $(1\pm\sign(N))/2$. Now we only have to specify 
whether the expression we have is for even $N$ of for odd $N$ and there is 
a one to one correspondence again. This is fully in line with the theorem 
that knowledge of only the even moments or only the odd moments is 
sufficient to reconstruct the function of $x$.

The method to do the inverse Mellin transform is now simple, provided one 
has a simple algorithm to say which index field of the harmonic 
polylogarithms corresponds to which index field of the harmonic sums. Such 
an algorithm exists~\cite{Remiddi:1999ew}.

1: Given a HS, construct the single term with a HP with the highest weight 
and the denominator $1/(1\pm x)$. Subtract and add.

2: Do a Mellin transform on the subtracted term. This will create a term 
that cancels the original HS and gives also terms with lower weight HS's.

3: Keep repeating this procedure untill there are no more terms with HS's.
The remaining constant terms will get a factor $\delta(1-x)$.

\section{Difference equations}

When we study diagrams as a function of $N$, usually one cannot come up 
with a simple expression in a direct constructive way. Instead one can 
construct equations of the type
\begin{eqnarray}
	A_0(N) F(N) + A_1(N) F(N\minus 1) + \nonumber \\
		\ \ \ \ \ \ \ \cdots + A_n(N) F(N\minus n) + G(N) & = & 0
\end{eqnarray}
Such an equation we call an n-th order difference equation. A simple 
example would be a first order equation which is also called a recursion. 
In that case the solution can be written down as a single sum:
\begin{eqnarray}
\label{eq:firstsol}
        F(N) & = & \sign(N)\frac{\prod_{j=1}^N A_1(j)}{\prod_{j=1}^N A_0(j)}F(0)
		\nonumber \\ & + &
                \sum_{i=1}^N\sign(N-i)\frac{\prod_{j=i+1}^N A_1(j)}{
			\prod_{j=i}^N A_0(j)}G(i)\, .
\end{eqnarray}
Under the right conditions the products can be written as combinations of 
$\Gamma$-functions in $N,i,j$ and $\epsilon$.

Recently other authors have also obtained difference equations for the 
evaluation of diagrams. Tarasov~\cite{Tarasov} obtains equations in the 
dimension of space-time. Laporta~\cite{Laporta} has a very promising and 
systematic approach that has similarities with the above method, except for 
that it does not involve the Mellin parameter $N$. This results in a 
completely different approach to solving them.

When we define basic building blocks (BBB) as diagrams in which the 
momentum $P$ of the parton flows only through a single line in the diagram, 
we have to worry about two BBB's when we do a two loop calculation and 17 
BBB's when we do a three loop calculation. Of these 17 seven are two loop 
diagrams with a one loop insertion. The two BBB's at the two loop level 
result in first order difference equations. This was already noticed 
in~\cite{Kazakov:1988jk}. The 17 BBB's at the three loop 
level split into 4 first order equations, 12 second order equations and one 
third order equation. We have been able to set up all these equations. It 
is possible to define a hierarchy amoung the integrals in which a more 
complicated diagram $F$ is related in this way to simpler diagrams inside 
the function $G$. For example the equation for a ladder diagram may have 
two loop diagrams with a one loop insertion inside the function $G$. The 
coefficients $A_i$ are usually polynomials in $N$ and $\epsilon = 2-D/2$. 
Let us assume that the function $G$ can be expressed in terms of harmonic 
sums. For a number of types of these equations one can then make a 
constructive solution for the function $F$. This is usually rather 
complicated and moreover, we did not find constructive methods for all the 
equations that we derived. There does however exist a quite good method 
that does the job for all equations. This method goes as follows:
\begin{itemize}
\item Divide the equation by $N^q$ in which q is the highest power of $N$ 
occurring in the coefficients $A_i$.
\item Synchronize the arguments of the sums in $G$ and the denominators.
\item Write
$S_{\vec{m}}(N)/N^p = S_{p,\vec{m}}(N)-S_{p,\vec{m}}(N-1)$ and similar 
forms when $\sign(N)$ is involved.
\item Assume that $F(N) = \sum c_{i,n,\vec{m}} \epsilon^i 
S_{\vec{m}}(N-n)$. We have to include powers of $\zeta_3$, $\zeta_4$ and 
$\zeta_5$ as well, but let us ignore that here. We have to guess a bit what 
range of values we need for $i$ and $n$.
\item We also bring the $A_j(N) F(N-j)$ terms into the standard form with 
just a single $S$-sum without denominators.
\item This defines a large number of equations in the $c_{i,n,\vec{m}}$ 
which we solve. If the range in $i$ and $n$ was large enough we should find 
a solution. The maximum weight we use defines (amoung others) to what power 
in $\epsilon$ our answer can be correct.
\item There will still be a few free parameters left, as we have not fixed 
the boundary conditions yet. For an $n$-th order equation we will need $n$ 
values. These can be computed with the Mincer program, as they concern 
fixed values of $N$.
\item The boundary conditions give a few more equations, and these fix all 
remaining constants. This then determines the function $F$.
\end{itemize}
This method is rather general and relatively fast. For the cases for which we 
could compare it with the constructive method it was significantly faster. 
It allows us to solve all equations we run into provided we do not run out 
of computer resources (in our method of solution there is a limitation in 
the 32 bits version of FORM~\cite{FORM} that limits the equations to between 1000 and 
2000 terms each. The 64 bit version has no realistic restriction in this 
respect. It would also be possible to choose a less efficient way of 
solving these equations if the need were to arise.)

Because we have to solve these equations all the way to weight 6 sums (of 
which there are 486 for each $n$) and for several values of $n$, these 
equations involve often a few thousand coefficients to be solved. 
Fortunately the system is rather sparse.

\section{Status}

We have used the equation solving program already to solve all equations 
that occur at the two loop level in which the two loop integral has all 
lines with integer powers, but one line has a power involving $\epsilon$ as 
is to be expected from three loop integrals of which one loop has been 
integrated already.
In addition we have constructed all equations for the basic building blocks 
at the three loop level. Because it is still much work to program the whole 
hierarchy we are still far away from completing the computation of the 
simplest function $G$, solving the next function $F$, computing the next 
$G$, then the next $F$, etc. all the way to the most complicated diagrams. 
The most complicated nonplanar basic building block is determined by a 
third order difference equation, while the most complicated benz type 
diagram has a second order equation that contains about 200000 terms before 
the integrals in the $G$-function are evaluated. After evaluation it should 
simplify considerably of course. There are however no more fundamental 
problems left to obtain solutions for the basic building blocks.

The reduction for composite diagrams into basic building blocks follows a 
similar scheme. Take for instance the diagram
\begin{eqnarray}
NO_{13} & = &
	\raisebox{-19.1pt}{
	\SetScale{0.4} \SetPFont{Helvetica}{20}
	\hspace{-15pt}
	\begin{picture}(100,50)(0,0)
	\SetColor{Blue}
	\SetWidth{1.2}
	\CArc(60,60)(40,120,270)
	\CArc(140,60)(40,270,60)
	\Line(60,20)(140,20)
	 \Line(60,100)(140,20)
	 \Line(140,100)(104,64)\Line(96,56)(60,20)
	\Line(0,60)(20,60)
	\Line(180,60)(200,60)
	\SetColor{Red}
	\SetWidth{4}
	\CArc(60,60)(40,90,120)
	\CArc(140,60)(40,60,90)
	\Line(60,100)(140,100)
	\SetColor{Black}
	\PText(20,90)(0)[r]{1}
	\PText(180,90)(0)[l]{1}
	\PText(20,30)(0)[r]{1}
	\PText(180,30)(0)[l]{1}
	\PText(65,70)(0)[l]{1}
	\PText(135,70)(0)[r]{1}
	\PText(100,15)(0)[t]{1}
	\PText(100,110)(0)[b]{1}
	\PText(50,105)(0)[b]{1}
	\PText(150,105)(0)[b]{1}
	\end{picture}
	} \nonumber \\ & = &
	\int dp_1^D dp_2^D dp_3^D
		\frac{1}{p_1^2 (P\plus p_1)^2 (P\plus p_2)^2}
				\nonumber \\ && \times \frac{1}{
				(P\plus p_3)^2 p_3^2p_4^2p_5^2p_6^2p_7^2p_8^2}
\end{eqnarray}
We derived a general formula for such diagrams that is relatively simple:
\begin{eqnarray}
	(N\plus a\plus b)\TCa(a,A,B,b) \ \ \ \ \ 
	& = \nonumber \\ \ \ \ \ \ \ a\TCa({a+1},{A-1},B,b)
		\nonumber \\ \ \ \ \ 
    + b\TCa(a,A,{B-1},{b+1}) \nonumber \\ \ \ \ \ 
    -(N\minus 1\plus n)\frac{2P\mydot Q}{Q\mydot Q}\TCa(a,A,B,b)
\end{eqnarray}
This allows us to express the $NO_{13}$ diagram into a simpler diagram by means 
of a simple recursion. Hence this can be solved, once we know the simpler 
diagram. We could either use the general program for difference equations 
for this, or write the solution as a single sum over this simpler diagram. 
This sum involves $\Gamma$-functions which have to be expanded (and give 
harmonic sums again) and in the end we can use some general procedures for 
such sums. The main problem is currently to derive all the equations and 
recursions that break all composite diagrams down into basic building 
blocks. This is just an enormous amount of work. After that all relations 
have to be programmed and tested which is another enormous amount of work.

\end{document}